\documentclass[aps,prb,twocolumn,groupedaddress,
showpacs,amsmath,amssymb]{revtex4}

\usepackage{graphicx}
\usepackage{dcolumn}
\usepackage{bm}
\begin{document}

\title{Achieving fast oxygen diffusion in perovskites by cation ordering}

\author{A.~A.~Taskin}
\email[]{kotaskin@criepi.denken.or.jp}
\author{A.~N.~Lavrov}
\altaffiliation{Present address: Institute of Inorganic Chemistry,
Novosibirsk 630090, Russia}
\author{Yoichi~Ando}
\affiliation{Central Research Institute of Electric Power Industry,
Komae, Tokyo 201-8511, Japan}


\begin{abstract}

The oxygen-exchange behavior has been studied in half-doped manganese
and cobalt perovskite oxides. We have found that the oxygen diffusivity
in Gd$_{0.5}$Ba$_{0.5}$MnO$_{3-\delta}$ can be enhanced by orders of
magnitude by inducing crystallographic ordering among lanthanide and
alkali-earth ions in the A-site sublattice. Transformation of a simple
cubic perovskite, with randomly occupied A-sites, into a layered crystal
GdBaMn$_2$O$_{5+x}$ (or isostructural GdBaCo$_2$O$_{5+x}$ for cobalt
oxide) with alternating lanthanide and alkali-earth planes reduces the
oxygen bonding strength and provides disorder-free channels for ion
motion, pointing to an efficient way to design new ionic conductors.

\end{abstract}


\maketitle

Oxygen ion conductors - solids exhibiting very fast oxygen diffusion -
constitute the basis for such emerging technologies as the membrane
oxygen separation or the solid-oxide fuel cell (SOFC) power generation.
\cite{1,2}. These technologies offer enormous economical and ecological
benefits provided high performance materials can be developed: The
scientific challenge is to design materials demonstrating high oxygen
diffusivity at low enough temperature.

In general, a crystal must meet two fundamental requirements to be a
good oxygen-ion conductor: (i) it must contain a lot of vacancies in the
oxygen sublattice, and (ii) the energy barrier for oxygen migration from
one site to another must be fairly small, typically less than $\sim
1$~eV. Only a few types of oxides, and perovskites ABO$_3$ (A is a
rare-earth or an alkali-earth element and B is typically a transition
metal) among them, have been found to possess these features.\cite{1, 2,
3, 4, 5, 6, 7, 8} Doped perovskites, which possess a high electronic
conductivity in addition to the ionic one, are considered for using as
electrodes in SOFC and as oxygen-selective membranes; for example,
strontium-doped lanthanum manganese oxide,
La$_{1-y}$Sr$_y$MnO$_{3-\delta}$, is a standard cathode material for
SOFC applications operating at temperatures around 1000$^{\circ}$C.
\cite{6} Recently, serious efforts are made to reduce the operation
temperature of SOFC, and for the operation at 700 - 800$^{\circ}$C,
strontium-doped lanthanum cobalt oxide,
La$_{1-y}$Sr$_y$CoO$_{3-\delta}$, is considered to be the most promising
cathode material. \cite{7,8} The performance of perovskite oxides has
been already optimized as much as possible mostly by means of various
ion substitutions in both A and B sublattices, \cite{6,7,8,9} but they
still fail to operate at low enough temperatures $\sim500$$^{\circ}$C
required for successful commercialization of the fuel cell technology.

In this Letter, we show that the oxygen-ion diffusion in a doped
perovskite can be enhanced by orders of magnitude if a simple cubic
crystal [schematically shown in Fig.1(a)] transforms into a layered
compound with ordered lanthanide and alkali-earth ions [Fig. 1(b)],
which reduces the oxygen bonding strength and provides disorder-free
channels for ion motion.

\begin{figure}
\includegraphics*[width=8.0cm]{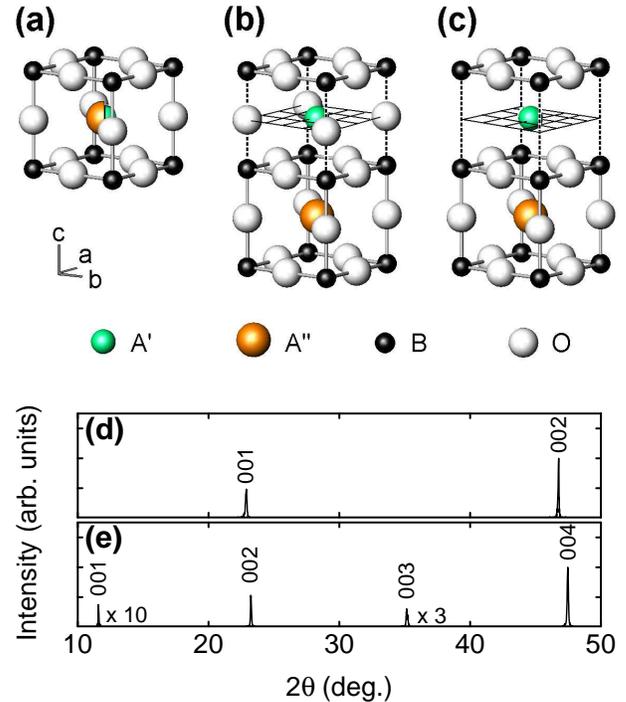}
\caption{Ordering of lanthanide A'$^{3+}$ and alkali-earth A''$^{2+}$
ions in the A-site sublattice of half-doped perovskites. (a) A simple
cubic perovskite A'$_{0.5}$A''$_{0.5}$BO$_3$ with random occupation of
A-sites is transformed into (b) a layered crystal A'A''B$_2$O$_6$ by
doubling the unit cell, provided the difference in ionic radii of A' and
A'' ions is sufficiently large. (c) Oxygen atoms can be partially or
completely removed from lanthanide planes in A'A''B$_2$O$_{5+x}$,
providing a variability of the oxygen content, $0\leq x \leq 1$. (d, e)
X-ray Bragg's (00L) peaks obtained for cubic
Gd$_{0.5}$Ba$_{0.5}$MnO$_{3-\delta}$ (d) and layered GdBaMn$_2$O$_{5+x}$
(e), which demonstrate the doubling of the unit cell along the $c$ axis
with the cation ordering (for convenience, the peak intensity is
multiplied by a factor indicated near each peak).}
\end{figure}

Recently, the A-site-ordered manganese and cobalt perovskite oxides have
been synthesized by several groups.\cite{10,11} We have succeeded in
growing high-quality single crystals of these compounds by the
floating-zone (FZ) technique.\cite{12} What makes these layered oxides
promising for searching a high oxygen mobility is their remarkable
variability of oxygen content: Oxygen atoms can be partially or even
completely removed from the lanthanide planes [as shown in Fig. 1 (c)],
creating a lot of vacant sites in the crystal lattice. While
polycrystalline air-sintered samples of GdBaMn$_2$O$_{5+x}$ adopt the
cubic perovskite crystal structure (and thus should be expressed as
Gd$_{0.5}$Ba$_{0.5}$MnO$_{3-\delta}$), the ordered phase can be grown
under strongly reducing atmosphere of pure argon. The fortunate
opportunity to obtain both layered GdBaMn$_2$O$_{5+x}$ and cubic
Gd$_{0.5}$Ba$_{0.5}$MnO$_{3-\delta}$ phases allows direct comparison of
the oxygen behaviour in materials with and without the A-site sublattice
ordering. For the cobalt oxide, on the other hand, only the ordered
GdBaCo$_2$O$_{5+x}$ phase can be obtained even at high oxygen pressure.

The homogeneity and cation stoichiometry of our crystals are confirmed
by the electron-probe microanalysis (EPMA) and the inductively-coupled
plasma (ICP) spectroscopy. Ordering of the Gd and Ba ions into
consecutive (001) layers is confirmed by the X-ray diffraction
measurements [Fig. 1(d) and (e)], which clearly demonstrate the doubling
of the unit cell along the $c$ axis for GdBaMn$_2$O$_{5+x}$ single
crystals. The behaviour of dc resistivity shows that both
GdBaMn$_2$O$_{5+x}$ and GdBaCo$_2$O$_{5+x}$ undergo an
insulator-to-metal transition upon increasing temperature; at high
temperatures, they possess a metallic conductivity $\sigma$ of the order
of $10^2$ and $10^3$~S/cm, respectively. These values are much larger
than any ionic conductivity available in solids, indicating that these
compounds should be mixed ionic and electronic conductors (MIECs).

It is well known that in MIECs, the ionic conductivity $\sigma_i$, which
is typically much smaller than the electronic conductivity $\sigma_e$,
may be determined from diffusion measurements.\cite{13,14} In oxides
with variable oxygen stoichiometry, like GdBaMn$_2$O$_{5+x}$ and
GdBaCo$_2$O$_{5+x}$, a sharp change in conditions (oxygen pressure or
temperature) results in a gradual transition into a new equilibrium
state that differs in oxygen content; this means that the oxygen
concentration in crystal samples changes with time, which can be
measured as a weight change. The observed relaxation time depends on two
processes: the oxygen exchange at the interface between the gas and the
solid, and the oxygen bulk diffusion.\cite{13,14} In the present study,
both the surface exchange coefficient $K$ and the chemical diffusion
coefficient $D$ are obtained by examining the weight change of crystals
as a function of time.

The oxygen-exchange behaviour in cubic manganese oxide and its layered
counterpart, studied in polycrystalline samples with comparable
morphology, is shown in Fig. 2. There are two clear consequences of the
A-site ordering: First, as shown in Fig. 2(a), the change in the
equilibrium oxygen content $x_{\text{eq}}$ upon decreasing oxygen
pressure from 1 to $10^{-5}$ bar at 700$^{\circ}$C is much more
pronounced in the layered oxide. Thus, it is much easier to remove
oxygen from the layered GdBaMn$_2$O$_{5+x}$ than from cubic
Gd$_{0.5}$Ba$_{0.5}$MnO$_{3-\delta}$. Second, and more important, is a
huge difference in the rate of the oxygen uptake. Figure 2(b) shows a
normalized change in the oxygen content $\Delta
x_{norm}=(x-x_0)/(x_{\text{eq}}-x_0)$ ($x_0$ is the initial oxygen
content and $x_{\text{eq}}$ is the equilibrium one), measured during the
annealing in the oxygen gas flow at 350$^{\circ}$C and 650$^{\circ}$C,
for samples with ordered (open symbols) and disordered (filled symbols)
A-site sublattice. The relaxation time is about two orders of magnitude
smaller in the layered oxide. Surprisingly, the oxygen accumulation in
the layered GdBaMn$_2$O$_{5+x}$ even at 350$^{\circ}$C turns out to be
faster than that in the cubic Gd$_{0.5}$Ba$_{0.5}$MnO$_{3-\delta}$ at
650$^{\circ}$C.

\begin{figure}
\includegraphics*[width=8.5cm]{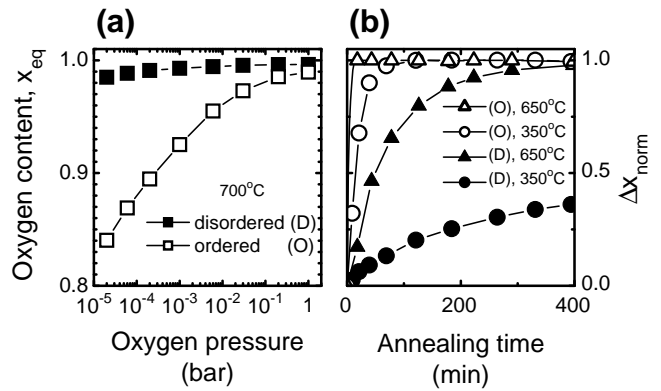}
\caption{Comparison of oxygen behaviour in cubic
Gd$_{0.5}$Ba$_{0.5}$MnO$_{3-\delta}$ with disordered A-site sublattice
(i.e. randomly distributed Gd$^{3+}$ and Ba$^{2+}$ ions) and layered
GdBaMn$_2$O$_{5+x}$. (a) Equilibrium oxygen content x$_{\text{eq}}$ (for
Gd$_{0.5}$Ba$_{0.5}$MnO$_{3-\delta}$, $x$=1-2$\delta$) measured at
700$^{\circ}$C as a function of the oxygen partial pressure. (b)
Normalized change in the oxygen content with time during annealing in
the oxygen gas flow at 350$^{\circ}$C and 650$^{\circ}$C.}
\end{figure}

Having established that the ordering of the A-site sublattice into a
layered structure significantly enhances the oxygen relaxation rate, we
now turn to a more quantitative analysis of the oxygen diffusion. For
that purpose, however, large single-crystal samples of
GdBaMn$_2$O$_{5+x}$ appear to be unsuitable; depending on $x$, the
orthorhombicity $(b-a)/a$ in GdBaMn$_2$O$_{5+x}$ varies from 0 up to
7\%, and the enormous strains emerging upon the oxygen intercalation
into large single crystals result in the formation of a regular array of
cracks on the crystal surface. We therefore focus our quantitative
analysis on another layered material GdBaCo$_2$O$_{5+x}$, which does not
have such a problem. To separate the surface-exchange and bulk-diffusion
contributions to the overall oxygen exchange rate, we study the oxygen
kinetics in GdBaCo$_2$O$_{5+x}$ single crystals of different size: the
oxygen kinetics is mostly limited by the surface exchange for small-size
crystals, while for large-size crystals it is mostly limited by the bulk
diffusion. The change in weight with time in a rectangular-shape sample
of a given size has an analytical solution,\cite{14} which is a function
of both the chemical diffusion coefficient $D$ and the surface exchange
coefficient $K$. Figure 3 shows the experimental data (symbols) and
analytical fits (solid lines) of the change in the oxygen content with
time for GdBaCo$_2$O$_{5+x}$ single-crystal samples of different sizes
upon annealing in the oxygen flow at 350$^\circ$C. From this set of
data, the parameters $D$ and $K$ can be uniquely determined to be $D=3
\cdot 10^{-7}$~cm$^2$/s and $K=2 \cdot 10^{-6}$~cm/s. We note
that the oxygen kinetics does not show any detectable difference upon
varying the sample size along the $c$ axis in the range of
30-500~$\mu$m, implying an essentially two-dimensional character of the
oxygen diffusion in this layered material.

\begin{figure}[!t]
\leftskip12pt
\includegraphics*[width=7.2cm]{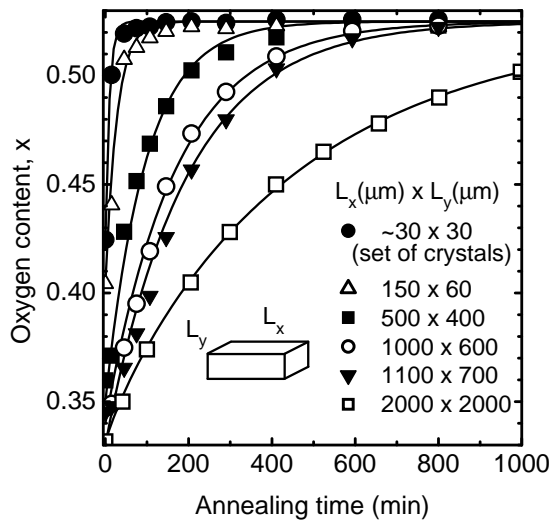}
\caption{Time dependence of the oxygen content in rectangular-shaped
GdBaCo$_2$O$_{5+x}$ single crystals of different sizes while annealing
in the oxygen flow at 350$^{\circ}$C (symbols). Solid lines represent
theoretical curves with a unique pair of parameters $D=3\cdot 10^{-7}$
cm$^2$/s and $K=2\cdot 10^{-6}$ cm/s for the whole set of experimental
data.}
\vspace{-3pt}
\end{figure}

Using the technique described above, we have determined the parameters
$D$ and $K$ for the temperature range of 250-600$^\circ$C at 1 bar
oxygen pressure [Fig. 4(a)]. These data are well fitted by the
activation laws $D($cm$^2/$s$)=0.15\,\cdot
\exp(-0.7\,$eV$/k_{\text{B}}T$) and $K($cm/s$)=15\,\cdot
\exp(-0.85\,$eV$/k_{\text{B}}T$), revealing quite low activation
energies for both processes. As a result, the oxygen diffusion in
GdBaCo$_2$O$_{5+x}$ becomes very fast already at rather low
temperatures, exceeding $10^{-5}$~cm$^2$/s at 600$^\circ$C. To
appreciate the true merit of this high oxygen diffusivity in
GdBaCo$_2$O$_{5+x}$, it is worth estimating the ionic conductivity,
which is uniquely determined by the self-diffusion (or trace-diffusion)
coefficient $D^*$.\cite{15} One should keep in mind that the chemical
diffusion coefficient $D$, obtained in our experiments, is related to
$D^*$ through a so-called thermodynamic factor $\Theta$ [$\equiv$
$(\partial \ln P/\partial \ln x)/2$], $D=\Theta \cdot D^*$, reflecting
the fact that the true driving force for diffusion is a gradient in the
chemical potential but not a gradient in the oxygen
concentration.\cite{15} In GdBaCo$_2$O$_{5+x}$, $\Theta$ turns out to be
quite low and does not exceed $\sim 10$ in the whole temperature and
oxygen-pressure range studied [Fig. 4(b)]. Thus, the ionic conductivity
of 0.01 S/cm, which is considered as a criterion for SOFCs to be
valuable\cite{1}, can be achieved in GdBaCo$_2$O$_{5+x}$ already at
$\sim 500$$^\circ$C. Clearly, there should still be a possibility for
further improvement of properties in ordered perovskite oxides, and the
most obvious way is a substitution of different rare earths for Gd.

\begin{figure}[h]
\vspace{7pt}
\includegraphics*[width=8.4cm]{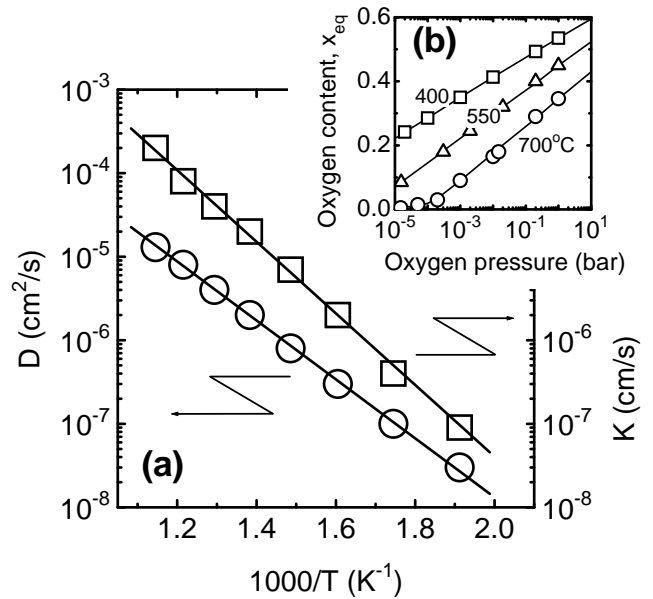}
\caption{Oxygen bulk diffusion and surface oxygen exchange in
GdBaCo$_2$O$_{5+x}$. (a) Temperature dependences of the diffusion
coefficient $D(T)$ and the surface exchange coefficient $K(T)$, obtained
from the oxygen kinetics experiments similar to those shown in Fig. 3.
The inset (b) shows the dependence of the equilibrium oxygen content
$x_{\text{eq}}$ on the oxygen pressure $P$ for several temperatures.}
\vspace{-10pt}
\end{figure}

In conclusion, we have found a remarkable enhancement of the oxygen
diffusivity in ordered perovskite oxides as exemplified in
GdBaMn$_2$O$_{5+x}$ and GdBaCo$_2$O$_{5+x}$, which opens a possibility
to develop a new class of materials suitable for applications that
require a fast oxygen transport in the intermediate temperature range.
The improvement of the oxygen transport properties induced by the cation
ordering in half-doped perovskites provides a good example of how the
formation of a layered crystal structure, with disorder-free channels
for ions migration and with a weakened bonding strength of oxygen, can
significantly facilitate the oxygen motion in the crystal lattice.



\begin{thebibliography}{99}

\bibitem{1} B. C. H. Steele and A. Heinzel,
Nature {\bf 414}, 345 (2001).

\bibitem{2} N. Q. Minh and T. Takahashi, 
{\it Science and Technology of Ceramic Fuel Cells} 
(Elsevier,   Amsterdam, 1995).

\bibitem{3} P. Lacorre, F. Goutenoire, O. Bohnke, R. Retoux, and
Y. Laligant, Nature {\bf 404}, 856 (2000).

\bibitem{4} J.B. Goodenough,
Annu. Rev. Mater. Res. {\bf 33}, 91 (2003).

\bibitem{5} Z. Shao and S.M. Haile,
Nature {\bf 431}, 170 (2004).

\bibitem{6} S.J. Skinner,
Int. J. Inorg. Mater. {\bf 3}, 113 (2001).

\bibitem{7} R.A. De Souza and J.A. Kilner,
Solid State Ionics {\bf 106}, 175 (1998).

\bibitem{8} J.A. Kilner,
Solid State Ionics {\bf 129}, 13 (2000).

\bibitem{9} Y.L. Yang, A.J. Jacobson, C.L. Chen, G.P. Luo,
K.D. Ross, and C.W. Chu,
Appl. Phys.   Lett., {\bf 79}, 776 (2001).

\bibitem{10} F. Millange, V. Caignaert, B. Domenges,
B. Raveau, and E. Suard,
Chem. Mater. {\bf 10}, 1974 (1998).

\bibitem{11} A. Maignan, C. Martin, D. Pelloquin,
N. Nguyen, and B. Raveau,
J. Solid State Chem. {\bf 142}, 247 (1999).

\bibitem{12} A.A. Taskin, A.N. Lavrov, and Y. Ando,
Phys. Rev. Lett. {\bf 90}, 227201 (2003).

\bibitem{13} I. Yasuda and M. Hishinuma,
J. Solid State Chem. {\bf 123}, 382 (1996).

\bibitem{14} J.A. Lane and J.A. Kilner,
Solid State Ionics {\bf 136-137}, 997 (2000).

\bibitem{15} A. Van der Ven, G. Ceder, M. Asta, and P.D. Tepesch,
Phys. Rev. B {\bf 64}, 184307  (2001).

\end{thebibliography}
\end{document}